\documentclass[12pt]{article}
\usepackage{subfloat}
\pdfoutput = 1
\usepackage{graphics}
\usepackage{graphicx} 
\usepackage{multirow}
\usepackage{hhline}
\begin{document}
\date{Today}
\title{{\bf{\Large  Holographic insulator/superconductor phase transition in higher dimensional Gauss-Bonnet gravity }}}

\author{ {\bf {\normalsize Diganta Parai}$^{a}$
\thanks{digantaparai007@gmail.com}},\,
{\bf {\normalsize Sunandan Gangopadhyay}$^{b}
$\thanks{sunandan.gangopadhyay@gmail.com, sunandan.gangopadhyay@bose.res.in}},\,
{\bf {\normalsize Debabrata Ghorai}$^{b}$
\thanks{debanuphy123@gmail.com, debabrataghorai@bose.res.in}}\\
$^{a}$ {\normalsize Indian Institute of Science Education and Research Kolkata}\\{\normalsize Mohanpur, Nadia 741246, India}\\[0.2cm] 
$^{b}$ {\normalsize  Department of Theoretical Sciences,}\\{\normalsize  S.N. Bose National Centre for Basic Sciences,}\\{\normalsize JD Block, 
Sector III, Salt Lake, Kolkata 700106, India}\\[0.2cm]
}
\date{}

\maketitle

\begin{abstract}
{\noindent In this work, we have analytically investigated the insulator/superconductor phase transition in the presence of $d$-dimensional Gauss-Bonnet AdS soliton background. Using the Sturm-Liouville eigenvalue method, we have calculated the value of the critical chemical potential $\mu_c$ in any arbitrary dimension $d\geq 5$. We have then studied the condensation operator values and charge density in terms of the chemical potential and discussed the $d=5, 6, 7$ cases using our general results in $d$ dimensions. Our analytical results agree very well with the numerically findings in the literature.}
\end{abstract}
\vskip 1cm

\section{Introduction}


\noindent Strongly coupled systems are known to be elusive since conventional perturbative approaches are not suitable to describe such systems. They are nevertheless ubiquitous in nature, and occurs especially in condensed matter physics. 
The profound discovery of the anti-de Sitter/conformal field theory (AdS/CFT) correspondence born from string theory as a possible theory of quantum gravity, has proved to be a robust theoretical framework to understand strongly coupled field theories through a weakly coupled gravitational system. It provides an exact correspondence between a gravitational theory in a $(d+1)$-dimensional AdS spacetime and a conformal field theory (CFT) sitting on the $d$-dimensional boundary of this spacetime \cite{adscft1}-\cite{adscft3}. 
The duality revealed that the asymptoticaly AdS black hole spacetime in the bulk can become unstable leading to the condensation of scalar hair below a certain critical temperature. This instability corresponds to a second order phase transition from normal to superconducting state thereby giving birth to the model of the holographic $s$-wave superconductor and owes its origin to the breaking of a local $U(1)$
symmetry near the event horizon of the black hole. A number of investigations have been made in this direction in order to understand various properties of holographic superconductor/metal phase transition in the framework of usual Maxwell electromagnetic theory \cite{adscft4}-\cite{dg3} as well as in Born-Infeld electrodynamics \cite{hs19}-\cite{dg6} which is a non-linear theory of electrodynamics. 
Examples of black holes exhibiting $p$-wave superconductivity has also been observed \cite{pufu}-\cite{sgp}. 

In addition to the holographic model of the bulk AdS black hole spacetime, it has also been realized that there can be a holographic model in the bulk Gauss-Bonnet (GB) AdS soliton background which has the ability to describe an insulator/superconductor phase transition. In this case the AdS soliton background becomes unstable to form condensates of the scalar field which is then interpreted as a superconducting phase for the chemical potential $\mu > \mu_c$, $\mu_c$ being a critical chemical potential.
The first investigation of this model was carried out numerically \cite{nrt} and considered 5-dimensional AdS soliton background. Thereafter, analytic investigations on this type of phase transition were carried out in \cite{cai-so}-\cite{lee}. The analytical technique that has been largely employed in studying these models have been the 
variational approach for the Sturm-Liouville (SL) eigenvalue problem. It has been observed that this analytic method is more efficient than the matching technique first discussed in \cite{gregory}. 

In the present work, we have investigated analytically a holographic model of insulator to superconductor phase transition in the AdS$_d$ soliton background and then study the $d=5, 6, 7$ cases. The motivation of carrying out this investigation stems from the fact that it would be interesting to look at the effects of background spacetimes in higher dimensions on the properties of holographic superconductors. To start we first set up the $d$-dimensional AdS soliton background with the matter field coupled with Maxwell electrodynamics. Using the analytical approach, namely, the Sturm-Liouville eigenvalue method, we calculate the critical chemical potential ($\mu_c$) for different values 
of the spacetime dimension $d$, GB parameter $\tilde{\alpha}$ and mass of the matter (scalar) field $mL^2_{eff}$ chosen to be consistent with the Breitenloher-Freedman bound \cite{Breitenloher}. 
We then calculate the condensation operator values and the charge density in terms of the chemical potential.
It has been observed that higher values of the dimension of spacetime, curvature correction and scalar field mass make the condensate harder to form. Our analytical results agree with the existing numerical results in the literature for $d=5$.

The paper is organized as follows. In section 2, we mention the basic set up for the insulator/superconductor phase transition in $AdS_d$ soliton background. In section 3, we have analytically calculated the critical chemical potential for different values of $d, \tilde{\alpha}$ and $mL^2_{eff}$. We then study the condensation operator values and charge density in terms of the chemical potential in section 4. We conclude in section 5.

\section{Basic set up}
\noindent The model of a holographic insulator/superconductor phase transition  in the GB $AdS_{d}$ soliton background can be constructed by considering the following action
\begin{eqnarray}
S=\int d^{d}x \sqrt{-g}\left[\frac{1}{16\pi G_{d}}(R-2\Lambda+\alpha R_{GB}) -\frac{1}{4} F^{\mu \nu} F_{\mu \nu} - (D_{\mu}\psi)^{*} D^{\mu}\psi-m^2 \psi^{*}\psi \right]
\label{5}
\end{eqnarray}
where $R$ is the Ricci scalar, $R_{GB}=R^2-4R_{\mu\nu}R^{\mu\nu}+R_{\alpha \beta \gamma \delta }R^{\alpha \beta \gamma \delta }$ is the GB term, $\alpha$ is the GB parameter,  $\Lambda=-(d-1)(d-2)/(2L^2)$ is the cosmological constant, $F_{\mu \nu}=\partial_{\mu}A_{\nu}-\partial_{\nu}A_{\mu}$ is the Maxwell field strength tensor, $D_{\mu}\psi=\partial_{\mu}\psi-iqA_{\mu}\psi$ is the covariant derivative,  $A_{\mu}$ and $ \psi $ represent the gauge and the scalar fields. 

\noindent The line element of the $d$-dimensional $AdS$ soliton in GB gravity reads \cite{Cai}
\begin{eqnarray}
ds^2=-r^2 dt^2+\frac{1}{f(r)}dr^2+f(r)d{\varphi}^2+ r^2 dx_{i} dx^{i}
\label{1}
\end{eqnarray}
with
\begin{eqnarray}
f(r)=\frac{r^2}{2\widetilde{\alpha}}\Bigg[1-\sqrt{1-\frac{4\widetilde{\alpha}}{L^2}\bigg(1-\frac{r_{s}^{d-1}}{r^{d-1}}\bigg)}\Bigg]
\label{2}
\end{eqnarray}
where $dx_{i}dx^{i}$ represents the line element of the $(d-3)$-dimensional hypersurface with no curvature, $r_{s}$ is the tip of the soliton, $L$ is the $AdS$ radius and $\widetilde{\alpha} $ is related to the GB coupling constant $\alpha$ as $\widetilde{\alpha}=(d-3)(d-4)\alpha $.
The interesting point about this geometry is that this does not have any horizon but a conical singularity at $r=r_{s}$. This singularity can be removed by imposing a period $\beta=\frac{4\pi L^2}{(d-1) r_{s}}$ for the coordinate $\varphi$. In the asymptotic region $(r\rightarrow \infty)$, $f(r)$ behaves as
\begin{eqnarray}
f(r)\sim\frac{r^2}{2\widetilde{\alpha}}\Bigg[1-\sqrt{1-\frac{4\widetilde{\alpha}}{L^2}}\Bigg].
\label{3}
\end{eqnarray}
Hence the effective asymptotic $AdS$ scale can be defined by \cite{nrt}
\begin{eqnarray}
L_{eff}^2=\frac{2\widetilde{\alpha}}{1-\sqrt{1-\frac{4\widetilde{\alpha}}{L^2}}}~.
\label{4}
\end{eqnarray}
\noindent It should be noted that $L_{eff}^{2}= L^{2}$ for $\widetilde{\alpha}\rightarrow 0 $ and $L_{eff}^{2}= \frac{L^{2}}{2}$ for $\widetilde{\alpha}\rightarrow \frac{L^2}{4} $. 
The Schwarzschild AdS soliton is recovered by taking the limit $\widetilde{\alpha}\rightarrow 0$ in eq.(\ref{2}).

\noindent Making the ansatz  
\begin{eqnarray}
\psi=\psi(r) ~,~~~~~ A_{\mu}=\phi(r)dt
\end{eqnarray}
 the equations of motion for the scalar field $\psi$ and gauge field $\phi$ read
\begin{eqnarray}
\psi^{\prime\prime}(r) + \left(\frac{f^{\prime}(r)}{f(r)}+\frac{d-2}{r}\right)\psi^{\prime}(r) + \left(\frac{q^{2}\phi^{2}(r)}{r^{2}f(r)}- \frac{m^{2}}{f(r)}\right)\psi(r) = 0
\label{7} \\
\phi^{\prime\prime}(r) + \left(\frac{f^{\prime}(r)}{f(r)}+\frac{d-4}{r}\right)\phi^{\prime}(r) - \frac{2 q^{2} \psi^{2}(r)}{f(r)}\phi(r) = 0 ~.
\label{8}
\end{eqnarray}
By introducing a new coordinate $z=r_{s}/r$, the above equations take the form
\begin{eqnarray}
\psi^{\prime \prime}(z) + \left(\frac{g^{\prime}(z)}{g(z)} - \frac{d-4}{z}\right)\psi^{\prime}(z) - \frac{1}{z^{4}g(z)} \Bigg[m^{2}-\left(\frac{q z\phi(z)}{r_{s}}\right)^2\Bigg]\psi(z)=0
\label{9} \\
\phi^{\prime \prime}(z) + \left(\frac{g^{\prime}(z)}{g(z)}-\frac{d-6}{z}\right) \phi^{\prime}(z)- \frac{2q^{2} \psi^{2}(z)}{ z^4 g(z)}\phi(z)= 0~.
\label{10}
\end{eqnarray}
Also under the above transformation of coordinates, the spacetime metric (\ref{2}) becomes
\begin{eqnarray}
\label{hjk1}
f(z)&=&r_{s}^{2}g(z) \nonumber \\
g(z)&=&\frac{1}{2\widetilde{\alpha}z^2}\bigg[1-\sqrt{1-4\widetilde{\alpha}\left(1-z^{d-1}\right)}\bigg].
\end{eqnarray} 
The rescalings $\psi\rightarrow \psi/q$ and $\phi\rightarrow (r_{s}/q)\phi$ allows one to set $q=1$ and $r_{s}=1$. So we can replace $g(z)$ in eq.(s)(\ref{9}, \ref{10}) by $f(z)$. For the rest of analysis we shall set $L=1$ for simplicity.

\noindent In order to solve the non-linear coupled differential equations (\ref{9}, \ref{10}), one needs to impose a boundary condition at the tip $z=1 ~(r=r_{s})$ and one at the origin $z=0 ~(r=\infty)$. Around the tip we can make the expansion 
\begin{eqnarray}
\label{11}
\psi(z)&=&a_{1}+a_{2}(z-1)+a_{3}(z-1)^2+....\\
\phi(z)&=&b_{1}+b_{2}(z-1)+b_{3}(z-1)^2+....
\label{12}
\end{eqnarray}
with $\psi(z=1)=a_{1}$ and $\phi(z=1)=b_{1}$.
\noindent The asymptotic behaviour of the fields can be written as 
\begin{eqnarray}
\label{13}
\psi(z)&=& \psi_{-}z^{\Delta_{-}}+\psi_{+}z^{\Delta_{+}} \\
\phi(z) &=& \mu-\rho z^{d-3}
\label{14}
\end{eqnarray}
with
\begin{eqnarray}
\Delta_{\pm}=\frac{1}{2}\left\{(d-1)\pm\sqrt{(d-1)^2+4m^{2}L_{eff}^{2}}\right\} .
\label{15}
\end{eqnarray}
\noindent From the AdS/CFT dictionary, $\psi_{\pm}$ can be interpreted as the expectation value of the condensation operator $\mathcal{O}_{\pm}$ in the dual field theory at the boundary. In this paper, we shall focus on the boundary condition in which $\psi_{-}=0$ and $\psi_{+} \neq 0$. In principle, one can do same analysis with the boundary condition $\psi_{+}=0$ and $\psi_{-} \neq 0$.
The constants $\mu$ and $\rho$ have the standard interpretations as the chemical potential and the charge density of the dual field theory respectively.

\section{The critical chemical potential $\mu_{c}$}
\noindent It has been observed numerically that when the chemical potential $\mu$ exceeds a critical chemical potential $\mu_{c}$, the condensations of the operators will happen. This can be viewed as a superconductor phase. For lower chemical potential that is when $\mu<\mu_{c}$ , the scalar field is zero and this can be viewed as an insulator phase \cite{nrt}. Therefore, the critical chemical potential $\mu_{c}$ is the turning point of this holographic insulator/superconductor phase transition.
Since $\psi(z)=0$ at $\mu=\mu_{c}$, eq.(\ref{10}) reduces to 
\begin{eqnarray}
\phi^{\prime \prime}(z) + \left(\frac{f^{\prime}(z)}{f(z)}-\frac{d-6}{z}\right) \phi^{\prime}(z)=0.
\label{17}
\end{eqnarray}
Multiplying by the integrating factor $f(z)z^{-(d-6)}$, the above equation can be recast as
\begin{eqnarray}
\frac{d}{dz}\left\{ \phi^{\prime}(z) f(z) z^{-(d-6)}  \right\} &=& 0 \nonumber \\
\Rightarrow  \frac{\phi^{\prime}(z) f(z)}{z^{d-6}}   &=& constant ~.
\end{eqnarray}
Using the fact $f(z=1)= 0$, we find that the above $constant=0$. Hence the gauge field equation finally reduces to 
\begin{eqnarray}
\phi^{\prime}(z) &=& 0 \nonumber \\
\Rightarrow \phi(z) &=& C =\mu
\label{hjk2}
\end{eqnarray}
where the constant of integration $C$ gets fixed from the asymptotic behaviour of the field $\phi(z)$ (\ref{14}).

\noindent Substituting the solution of the gauge field $\phi(z)$ in eq.(\ref{9}), we obtain the matter field equation near the critical chemical potential $\mu\rightarrow\mu_{c}$ to be

\begin{eqnarray}
\psi^{\prime \prime}(z) + \left(\frac{f^{\prime}(z)}{f(z)} - \frac{d-4}{z}\right)\psi^{\prime}(z) + \frac{1}{z^{4}f(z)} \left( \mu^{2}z^{2}-m^{2}\right)\psi(z)=0 ~.
\label{18}
\end{eqnarray}
Near the boundary, the matter field is defined as \cite{hs8}
\begin{eqnarray}
\psi(z)\sim\langle{\mathcal{O}_{\pm}\rangle}z^{\Delta_{\pm}}F(z)
\label{19}
\end{eqnarray}
with the trial function $F(z)$ satisfying the boundary conditions $F(0)=1$ and $F^{\prime}(0) =0$. Substituting eq.(\ref{19}) in eq.(\ref{18}), we obtain 
\begin{eqnarray}
F^{\prime\prime}(z)+\Bigg\{\frac{2\Delta_{\pm}}{z}+\left( \frac{f^{\prime}(z)}{f(z)}-\frac{d-4}{z}\right)\Bigg\}F^{\prime}(z)+\Bigg\{\frac{\Delta_{\pm}(\Delta_{\pm}-1)}{z^2}\nonumber\\
+\frac{\Delta_{\pm}}{z}\bigg(\frac{f^{\prime}(z)}{f(z)}-\frac{d-4}{z}\bigg)\Bigg\}+ \frac{1}{z^{4}f(z)} \left( \mu^{2}z^{2}-m^{2}\right)F(z)=0 
\label{20}
\end{eqnarray}
to be solved subject to the boundary condition $F^{\prime}(0)=0$.

\noindent The above equation can be recast in the Sturm-Liouville form
\begin{eqnarray}
\frac{d}{dz}\big\{p(z)F^{\prime}(z)\big\}+q(z)F(z)+\mu^{2}r(z)F(z)=0
\label{21}
\end{eqnarray}
with
\begin{eqnarray}
&p(z)&=\frac{z^{2\Delta_{\pm}-d+2}}{2\widetilde{\alpha}}\bigg[\sqrt{1+4\widetilde{\alpha}\left(z^{d-1}-1\right)}-1\bigg]\nonumber\\
&q(z)&=p(z)\Bigg[\frac{\Delta_{\pm}(\Delta_{\pm}-1)}{z^2}+\frac{\Delta_{\pm}}{z}\Bigg\{ \frac{(d-5) z^{d-2}-2 \big[z f(z)-\frac{2}{z}\big]}{2 \left(z^{d-1}-1\right)+z^2f(z)}-\frac{d-4}{z}\Bigg\}-\frac{m^2}{z^{4}f(z)}\Bigg]\nonumber\\
&r(z)&=z^{2\Delta_{\pm}-d+2} ~.
\label{22}
\end{eqnarray}
To estimate the eigenvalue $\mu^2$, we write down an expression for $\mu^2$, extremization of which leads to eq.(\ref{21}). \\
\noindent This reads 
\begin{eqnarray}
\mu^{2}=\frac{\int_{0}^{1}dz\big\{p(z)[F^{\prime}(z)]^{2}-q(z)[F(z)]^{2}\big\}}{\int_{0}^{1}dz r(z)[F(z)]^{2}}~.
\label{23}
\end{eqnarray}
We now use the trial function 
\begin{eqnarray}
F(z)=1-a z^2
\label{24}
\end{eqnarray}
which satisfies the boundary conditions $F(0)=1$ and $F^{\prime}(0)=0$, with $a$ being a constant. Substituting eq.(s)(\ref{22}),(\ref{24}) in eq.(\ref{23}), we get


\begin{eqnarray}
\mu^2=\frac{U(\widetilde{\alpha},d,m)-V(\widetilde{\alpha},d,m)a+W(\widetilde{\alpha},d,m)a^2}{\frac{1}{2\Delta-d+3}-\frac{2a}{2\Delta-d+5}+\frac{a^2}{2\Delta-d+7}}
\label{25}
\end{eqnarray}
where
\begin{eqnarray}
U(\widetilde{\alpha},d,m)&=&\frac{1}{2\Delta-d+1}\Bigg[m^2+\frac{\Delta^2}{2\widetilde{\alpha}}\bigg\{1-\left(\sqrt{1-4\widetilde{\alpha}}\right)\, _2F_1\left(-\frac{1}{2},\frac{2\Delta-d+1}{d-1};\frac{2\Delta}{d-1};\frac{4 \widetilde{\alpha}}{4 \widetilde{\alpha}-1}\right)\bigg\}\Bigg]\nonumber\\
V(\widetilde{\alpha},d,m)&=&\frac{2}{2\Delta-d+3}\Bigg[m^{2}+\frac{\Delta ^2}{2\widetilde{\alpha}}\bigg\{1-\left(\sqrt{1-4 \widetilde{\alpha}}\right) \, _2F_1\left(-\frac{1}{2},\frac{2\Delta-d+3}{d-1};\frac{2 (\Delta +1)}{d-1};\frac{4\widetilde{\alpha}}{4\widetilde{\alpha}-1}\right)\bigg\}\nonumber\\
&-&\frac{\Delta (2\Delta-d+1)}{2\widetilde{\alpha}}\bigg\{1-\left(\sqrt{1-4 \widetilde{\alpha}}\right) \, _2F_1\left(\frac{1}{2},\frac{2\Delta-d+3}{d-1};\frac{2 (\Delta +1)}{d-1};\frac{4\widetilde{\alpha}}{4\widetilde{\alpha}-1}\right)\nonumber\\
&-&\frac{(4\Delta-d+1)(2\Delta-d+3)}{(\Delta+1)(2\Delta-d+1)}\left(\frac{\widetilde{\alpha}}{\sqrt{1-4 \widetilde{\alpha}}}\right)\, _2F_1\left(\frac{1}{2},\frac{2 (\Delta +1)}{d-1};\frac{2\Delta +d+1}{d-1};\frac{4\widetilde{\alpha}}{4\widetilde{\alpha}-1}\right)\bigg\}\Bigg]\nonumber\\
W(\widetilde{\alpha},d,m)&=&\frac{1}{2\Delta-d+5}\Bigg[m^{2}+ \frac{\Delta ^{2}+4}{2\widetilde{\alpha}}-\frac{2\sqrt{1-4 \widetilde{\alpha}}}{\widetilde{\alpha}}\left(1+\frac{\Delta ^2}{4} \right ) \, _2F_1\left(-\frac{1}{2},\frac{2\Delta-d+5}{d-1};\frac{2 (\Delta +2)}{d-1};\frac{4\widetilde{\alpha}}{4\widetilde{\alpha}-1}\right)\nonumber\\
&-&\frac{\Delta (2\Delta-d+1)}{2\widetilde{\alpha}}\bigg\{1-\left(\sqrt{1-4 \widetilde{\alpha}} \right)\, _2F_1\left(\frac{1}{2},\frac{2\Delta-d+5}{d-1};\frac{2 (\Delta +2)}{d-1};\frac{4\widetilde{\alpha}}{4\widetilde{\alpha}-1}\right)\nonumber\\
&-&\frac{(2\Delta-d+5)(4\Delta-d+1)}{(\Delta+2)(2\Delta-d+1)}\frac{\widetilde{\alpha}}{\sqrt{1-4 \widetilde{\alpha}}}\, _2F_1\left(\frac{1}{2},\frac{2 (\Delta +2)}{d-1};\frac{2\Delta +d+3}{d-1};\frac{4\widetilde{\alpha}}{4\widetilde{\alpha}-1}\right)\bigg\}\Bigg].
\label{26}
\end{eqnarray}

\noindent For $d=5,mL_{eff}^2=0, \widetilde{\alpha}=0.0001$ and $\Delta=\Delta_{-}= 4$, we obtain from eq.(\ref{25})
\begin{eqnarray} 
\mu^{2}=\frac{ 20.0013-32.0018 a+15.0008 a^2}{1.66667-2.5 a+ a^2}~.
\end{eqnarray}
To estimate the critical chemical potential we now need to minimize $\mu$ with respect to $a$. The value of $\mu$ attains minimum value at $a=0.44$ and the value of $\mu_{c}$ is $3.407$ which agrees with the finding in \cite{qjb}.\\
\noindent For $d=6,mL_{eff}^2=0,\widetilde{\alpha}=0.1$ and $\Delta=5$, we find
\begin{eqnarray} 
\mu^{2}=\frac{29.6808-48.985 a+22.6861 a^2}{1.57143-2.44444 a+ a^2} 
\end{eqnarray}
which gives the minimum value of the critical chemical potential to be $\mu_{c}=4.257$ at $a=0.513$. \\
\noindent For different values of the dimension of spacetime, GB factor and the mass of the scalar field, we have presented the values of the critical chemical potential $\mu_{c}$ in Table \ref{ta1} by following the above procedure.
It can be observed that higher values of the dimension of spacetime, curvature correction and scalar field mass make the condensate difficult to form. Our findings match with the numerical findings \cite{nrt}.
\begin{table}[h!]
\caption{Analytical values of critical chemical potential $(\mu_{c})$ for different values of $\tilde{\alpha}, d$ and $mL^2_{eff}$ }   
\centering                          
\begin{tabular}{|c| c| c| c| c|}            
\hline                                 
$ d $ & $mL^2_{eff}$ & \multicolumn{3}{c|}{$\tilde{\alpha}$}  \\
\hhline{~~---}
& & 0.0001 & 0.1 & 0.2 \\
\hline
 & 0 & 3.407 & 3.559 & 3.790  \\ 
\hhline{~----}
5 & -2 & 2.817 & 2.937 & 3.106  \\ 
\hhline{~----}
 & -$\frac{15}{4}$ & 1.890 & 1.963 & 2.042  \\ 
\hline
 & 0 & 4.068 & 4.257 & 4.550   \\    
\hhline{~----}
6 & -2 & 3.625 & 3.79 & 4.036   \\
\hhline{~----}
 & -$\frac{15}{4}$ & 3.138 & 3.277 & 3.474   \\ 
\hline      
 & 0 & 4.713 & 4.938 & 5.292   \\    
\hhline{~----}
7 & -2 & 4.354 & 4.560 & 4.877   \\
\hhline{~----}
 & -$\frac{15}{4}$ & 4.00 & 4.18 & 4.960   \\ 
\hline      
\end{tabular}
\label{ta1}  
\end{table}


\section{Condensation operator and critical exponent}

\noindent To investigate the critical exponent of the condensation operator and the relations between the charge density and chemical potential, we look at the field eq.(\ref{10}) near the critical chemical potential $\mu_{c}$. Substituting eq.(\ref{19}) in eq.(\ref{10}), we get
\begin{eqnarray}
\phi^{\prime\prime}(z)+\bigg\{\frac{f^{\prime}(z)}{f(z)}-\frac{d-6}{z}\bigg\}\phi^{\prime}(z)=\frac{2\langle{\mathcal{O}_{\pm}\rangle^{2}z^{2\Delta_{\pm} -4}F^{2}(z)}}{f(z)}\phi(z)~.
\label{27}
\end{eqnarray}
Near the critical point, we can expand $\phi(z)$ in the small parameter $\langle\mathcal{O}_{\pm}\rangle$ as
\begin{eqnarray}
\phi(z) \sim \mu_{c}+\langle{\mathcal{O}_{\pm}\rangle}^{2}\chi(z)
\label{28}
\end{eqnarray}
with the boundary condition  $\chi(z=1)=0$ near the tip.
Substituting this form of $\phi(z)$ in eq.(\ref{27}) and comparing the coefficient of $\langle{\mathcal{O}_{\pm}\rangle}^{2}$ on both sides of the equation, we get the equation of motion for $\chi(z)$ to be
\begin{eqnarray}
\chi^{\prime\prime}(z)+\bigg\{\frac{f^{\prime}(z)}{f(z)}-\frac{d-6}{z}\bigg\}\chi^{\prime}(z)=\frac{2\mu_{c} z^{2\Delta_{\pm} -4}F^{2}(z)}{f(z)}~.
\label{29}
\end{eqnarray}
Multiplying both sides of the above equation by the integrating factor $T(z)$ given by 
\begin{eqnarray}
T(z)=\frac{\sqrt{1+4\widetilde{\alpha}(z^{d-1}-1)}-1}{2\widetilde{\alpha}z^{d-4}}
\label{30}
\end{eqnarray}
the equation for $\chi(z)$ takes the form
\begin{eqnarray}
\frac{\mathrm{d} }{\mathrm{d} z}\Bigg[T(z)\chi^{\prime}(z)\Bigg]+2\mu_{c}z^{2\Delta _{\pm}-d+2}F^{2}(z)=0.
\label{31}
\end{eqnarray}
Integrating eq.(\ref{31}) between the limits $z=1$ to $z$ and using $T(z=1)=0$ yields
\begin{eqnarray}
T(z)\chi^{\prime}(z)=-2\mu_{c}\int_{1}^{z}dx(1-a x^2)^{2}x^{2\Delta _{\pm}-d+2}~.
\label{32}
\end{eqnarray}
Again integrating both sides with respect to $z$ from 0 to 1 , we obtain
\begin{eqnarray}
\chi(0)=-2\mu_{c}\int_{0}^{1}dz \frac{z^{d-6}}{f(z)}\bigg[\int_{1}^{z}dx(1-a x^2)^{2}x^{2\Delta _{\pm}-d+2}\bigg]~.
\label{33}
\end{eqnarray}
Now $\phi(z)$ near $z=0$ can be expanded as
\begin{eqnarray}
\phi(z) \sim \mu-\rho z^{d-3}=\mu_{c}+\langle{\mathcal{O}_{\pm}\rangle}^{2}\bigg\{\chi(0)+z\chi^{\prime}(0)+...+\frac{z^{d-3}}{(d-3)!}\chi^{d-3}(0)+...\bigg\}.
\label{34}
\end{eqnarray}
Comparing the coefficients of $z$ on both sides of the above equation, we obtain
\begin{eqnarray}
\mu-\mu_{c}=\langle{\mathcal{O}_{\pm}\rangle}^{2}\chi(0)
\label{35}
\end{eqnarray}
\begin{eqnarray}
-\rho=\langle{\mathcal{O}_{\pm}\rangle}^{2}\frac{\chi^{d-3}(0)}{(d-3)!}
\label{36}
\end{eqnarray}
together with $\chi^{\prime}(0)=\chi^{\prime\prime}(0)=......=\chi^{(d-4)}(0)=0$.\\
\noindent From eq.(s)({\ref{33}}), ({\ref{35}}), we can deduce
\begin{eqnarray}
\langle{\mathcal{O}_{\pm}\rangle}=\frac{\sqrt{\mu-\mu_{c}}}{\Bigg[4\widetilde{\alpha}\mu_{c}\int_{0}^{1}dz \frac{z^{d-4}\big\{\int_{1}^{z}dx(1-a x^2)^{2}x^{2\Delta _{\pm}-d+2}\big\}}{\big\{\sqrt{1+4\widetilde{\alpha}\left(z^{d-1}-1\right)}-1\big\}}\Bigg]^\frac{1}{2}}~.
\label{37}
\end{eqnarray}
From the above result we find that the critical exponent of
the condensation operator is equal to $1/2$.

\noindent Integrating both sides of eq.(\ref{31}) from $z=0$ to 1, we get
\begin{eqnarray}
\frac{1}{L_{eff}^{2}}\Bigg[\frac{\chi^{\prime}(z)}{z^{d-4}}\Bigg]_{z\rightarrow 0}
&=& -2\mu_{c}\int_{0}^{1}dz(1-a z^2)^{2}z^{2\Delta _{\pm}-d+2}~.
\label{38}
\end{eqnarray}
Now note that $\chi^{\prime}(z)$ and $(d-3)^{th}$ derivative of $\chi(z)$ at $z=0$ are related by 
\begin{eqnarray}
\Bigg[\frac{\chi^{\prime}(z)}{z^{d-4}}\Bigg]_{z\rightarrow 0}=\frac{\chi^{d-3}(0)}{(d-4)!} ~, ~~~~~~~d>4 .
\label{39}
\end{eqnarray}
Using this and from eq.(s)(\ref{36}, \ref{38}), we get the relation between the charge density and the condensation operator to be
\begin{eqnarray}
\rho&=&-\frac{\langle{\mathcal{O}_{\pm}\rangle}^2}{d-3}\Bigg[\frac{\chi^{\prime}(z)}{z^{d-4}}\Bigg]_{z\rightarrow 0}\nonumber\\
&=&\frac{2\mu_{c}L_{eff}^2\langle{\mathcal{O}_{\pm}\rangle}^{2}}{(d-3)}\int_{0}^{1}dz(1-a z^2)^{2}z^{2\Delta _{\pm}-d+2}~.
\label{40}
\end{eqnarray}
Plugging the expression for $\langle{\mathcal{O}_{\pm}\rangle}$ from eq.(\ref{37}) in eq.(\ref{40}), we obtain
\begin{eqnarray}
\rho=\Gamma (\widetilde{\alpha},d,m)(\mu-\mu_{c})
\label{41}
\end{eqnarray}
where
\begin{eqnarray}
\Gamma (\widetilde{\alpha},d,m)=\frac{\int_{0}^{1}dz(1-a z^2)^{2}z^{2\Delta _{\pm}-d+2}}{(d-3)(1-\sqrt{1-4\widetilde{\alpha}})\int_{0}^{1}dz \frac{z^{d-4}\big\{\int_{1}^{z}dx(1-a x^2)^{2}x^{2\Delta _{\pm}-d+2}\big\}}{\big\{\sqrt{1+4\widetilde{\alpha}\left(z^{d-1}-1\right)}-1\big\}}}~.
\label{42}
\end{eqnarray}
The above results give the general results valid in any spacetime timension $d\geq 5$. For $d=5$ and $mL^2_{eff}=-\frac{15}{4}$, the condensation operator values and the charge density read $\langle \mathcal{O}_{+}\rangle = 1.801 \sqrt{\mu - \mu_{c}}$ and $\rho= 1.330 (\mu-\mu_{c})$ which agree with the findings in \cite{Cai}. 

\section{Conclusion}
We now summarize our findings in this work. Using the Sturm-Liouville eigenvalue method, we have investigated analytically a holographic insulator/superconductor phase transition in the Gauss-Bonnet gravity in the AdS$_d$ soliton background. The set up that we have considered is that of a $d$-dimensional AdS soliton background with the matter field coupled with Maxwell electrodynamics. 
For $\mu < \mu_c$, the AdS soliton background is stable and the dual field theory 
can be interpreted as an insulator whereas for $\mu > \mu_c$, the AdS soliton background will be unstable to forming condensates of the scalar field which is interpreted as the superconducting phase in the dual field theory. Using this basic idea, we have calculated the critical chemical potential ($\mu_c$) for different values of $d, \tilde{\alpha}$ and $mL^2_{eff}$ 
for the case $\psi_{+}\neq 0$ and $\psi_{-}=0$ which are displayed in Table \ref{ta1}. Furthermore, we have calculated the condensation operators values and charge density in term of the chemical potential. Our analytical results agree with the numerical findings \cite{nrt} and previous analytical findings \cite{Cai} (for $\tilde{\alpha}=0$ case). The results show that the condensate gets to form harder for higher value of GB parameters. It has been observed that higher values of the dimension of spacetime and scalar field mass make the condensate harder to form. Further, from the expression of the condensation operator value, we also find that the critical exponent of
the condensation operator is equal to $1/2$.

\section*{Acknowledgments}
DP would like to thank the CSIR for financial support. DG would like to thank DST-INSPIRE for financial support. S.G. acknowledges the support by DST SERB under Start Up Research Grant (Young Scientist), File No.YSS/2014/000180. SG also acknowledges the support under the Visiting Associateship programme of IUCAA, Pune.

\end{document}